\documentstyle[psfig]{mn}

\newif\ifAMStwofonts

\def\g{SN~1995G}
\def\z{SN~1988Z}
\def\w{SN~1994W}
\def\cy{SN~1997cy}
\def\kms{km s$^{-1}$}
\def\Ha{H$\alpha$}
\def\Hb{H$\beta$}

\def\m100{mag/100$^d$}

\def\c57{{$^{57}$Co}\/}

\def\ti44{{$^{44}$Ti}\/}
\def\r0{{$R_0$}}

\ifoldfss
  \ifCUPmtlplainloaded \else
    \NewTextAlphabet{textbfit} {cmbxti10} {}
    \NewTextAlphabet{textbfss} {cmssbx10} {}
    \NewMathAlphabet{mathbfit} {cmbxti10} {} 
    \NewMathAlphabet{mathbfss} {cmssbx10} {} 
  \fi
  \ifAMStwofonts
    \ifCUPmtlplainloaded \else
      \NewSymbolFont{upmath} {eurm10}
      \NewSymbolFont{AMSa} {msam10}
      \NewMathSymbol{\upi}     {0}{upmath}{19}
      \NewMathSymbol{\umu}     {0}{upmath}{16}
      \NewMathSymbol{\upartial}{0}{upmath}{40}
      \NewMathSymbol{\leqslant}{3}{AMSa}{36}
      \NewMathSymbol{\geqslant}{3}{AMSa}{3E}

      \let\geq=\geqslant \let\ge=\geqslant
    \fi
  \fi
\fi 

\ifnfssone
  \newmathalphabet{\mathit}
  \addtoversion{normal}{\mathit}{cmr}{m}{it}
  \addtoversion{bold}{\mathit}{cmr}{bx}{it}
  \newmathalphabet{\mathbfit} 
  \addtoversion{normal}{\mathbfit}{cmr}{bx}{it}
  \addtoversion{bold}{\mathbfit}{cmr}{bx}{it}
  \newmathalphabet{\mathbfss} 
  \addtoversion{normal}{\mathbfss}{cmss}{bx}{n}
  \addtoversion{bold}{\mathbfss}{cmss}{bx}{n}
  \ifAMStwofonts
    \ifCUPmtlplainloaded \else
      \UseAMStwoboldmath
      \makeatletter
      \new@mathgroup\upmath@group
      \define@mathgroup\mv@normal\upmath@group{eur}{m}{n}
      \define@mathgroup\mv@bold\upmath@group{eur}{b}{n}
      \edef\UPM{\hexnumber\upmath@group}
      \new@mathgroup\amsa@group
      \define@mathgroup\mv@normal\amsa@group{msa}{m}{n}
      \define@mathgroup\mv@bold\amsa@group{msa}{m}{n}
      \edef\AMSa{\hexnumber\amsa@group}
      \makeatother
      \mathchardef\upi="0\UPM19
      \mathchardef\umu="0\UPM16
      \mathchardef\upartial="0\UPM40
      \mathchardef\leqslant="3\AMSa36
      \mathchardef\geqslant="3\AMSa3E

      \let\geq=\geqslant \let\ge=\geqslant
    \fi
  \fi
\fi 

\ifnfsstwo
  \DeclareMathAlphabet{\mathbfit}{OT1}{cmr}{bx}{it}
  \SetMathAlphabet\mathbfit{bold}{OT1}{cmr}{bx}{it}
  \DeclareMathAlphabet{\mathbfss}{OT1}{cmss}{bx}{n}
  \SetMathAlphabet\mathbfss{bold}{OT1}{cmss}{bx}{n}
  \ifAMStwofonts
    \ifCUPmtlplainloaded \else
      \DeclareSymbolFont{UPM}{U}{eur}{m}{n}
      \SetSymbolFont{UPM}{bold}{U}{eur}{b}{n}
      \DeclareSymbolFont{AMSa}{U}{msa}{m}{n}
      \DeclareMathSymbol{\upi}{0}{UPM}{"19}
      \DeclareMathSymbol{\umu}{0}{UPM}{"16}
      \DeclareMathSymbol{\upartial}{0}{UPM}{"40}
      \DeclareMathSymbol{\leqslant}{3}{AMSa}{"36}
      \DeclareMathSymbol{\geqslant}{3}{AMSa}{"3E}

      \let\geq=\geqslant \let\ge=\geqslant
    \fi
  \fi
\fi 

\ifCUPmtlplainloaded \else
  \ifAMStwofonts \else 
    \def\upi{\pi}
    \def\umu{\mu}
    \def\upartial{\partial}
  \fi
\fi
\title[The Type IIn SN~1995G: Interaction with the CSM.]
  {The Type IIn SN 1995G: Interaction with the CSM. 
\thanks{Based on observations collected at ESO
  - La Silla (Chile), Asiago (Italy) and Lick (USA)}}
\author[Pastorello et al.]
        {Pastorello,A.$^{1}$, Turatto,M.$^{2}$, Benetti,S.$^{2}$,
         Cappellaro,E.$^{2}$, Danziger,I.J.$^{4}$, 
	         \and Mazzali,P.A.$^{3}$, Patat,F.$^{5}$, Filippenko,A.V.$^{6}$,
         Schlegel,D.J.$^{6,7}$, Matheson,T.$^{6,8}$\\
         $^{1}$Dipartimento di Astronomia, Universit\`a di Padova,
         vicolo dell'Osservatorio 2, 35122 Padova, Italy\\
         $^{2}$Osservatorio Astronomico di Padova, vicolo dell'Osservatorio 5,
         35122 Padova, Italy\\
         $^{3}$Osservatorio Astronomico di Capodimonte, via Moiariello 16,
         80131 Napoli, Italy\\
         $^{4}$Osservatorio Astronomico di Trieste, via G.B. Tiepolo 11,
         34131 Trieste, Italy\\
         $^{5}$European Southern Observatory, Karl Schwarzschild-Str. 2, 
         D-85748, Garching bei M\"unchen, Germany\\
         $^{6}$Department of Astronomy, University of California, Berkeley, 
         CA 94720--3411, USA\\
         $^{7}$Present address: Department of Astrophysical Sciences, 
         Peyton Hall, 
         Princeton University, Princeton, NJ 08544, USA\\
         $^{8}$Present address: Center for Astrophysics, 60 Garden Street, 
         Cambridge, MA 02138, USA}
\date{Accepted .....;
      Received ....;
      in original form ....}

\pubyear{2001}

\begin{document}

\maketitle

\label{firstpage}

\begin{abstract}

We present the photometric and spectroscopic evolution of the type IIn
\g\/ in NGC 1643, on the basis of 4 years of optical and infrared
observations.  This supernova shows very flat optical light curves
similar to SN 1988Z, with a slow decline rate at all times. The
spectra are characterized by strong Balmer lines with multiple
components in emission and with a P-Cygni absorption component
blueshifted by only 700 \kms.  This feature indicates the presence of
a slowly expanding shell above the SN ejecta as in the case of SNe
1994aj and 1996L.  As in other SNe IIn the slow luminosity decline
cannot be explained only with a radioactive energy input and an
additional source of energy is required, most likely that produced by
the interaction between supernova ejecta and a pre--existent
circumstellar medium.  It was estimated that the shell material has a
density $n_H >> 10^{8}$ cm$^{-3}$, consistent with the absence of
forbidden lines in the spectra.  About 2 years after the burst the
low velocity shell is largely overtaken by the SN ejecta and the luminosity 
drops at a faster rate.

\end{abstract}


\section{Introduction} 

Supernovae of type II (SNe II) are characterized by H lines in their
spectra.  Among them a sub-class has been isolated which shows slow
photometric evolution, absence of broad P-Cyg absorptions, and
emission lines (\Ha, in particular) with multiple components. Schlegel
(1990) labeled these SNe as type IIn, where the ``n'' stands
for``narrow'', though this designation is somewhat misleading because
of the simultaneous presence of narrow and broad components. Best
studied cases of this class are \z\/ (Stathakis \& Sadler, 1991;
Turatto et al., 1993; Filippenko, 1997; Aretxaga et al., 1999) and
SN~1995N (Fox et al., 2000; Fransson et al., 2001). We recall that some
very bright SNe IIn with very broad emission components, implying velocities
as large as 17000 \kms, may be associated with GRB, namely SN~1997cy
\cite{tura00} and SN~1999E \cite{fili99,cap99a}. 
Another group, sometimes called SNe IId (or SNe IIsw), shows narrow
P--Cygni absorptions on top of otherwise normal SNe II spectra (SN~1984E
\cite{dopi84}, SN~1994aj \cite{bene98}, SN~1996L \cite{bene99} and
SN~1996al \cite{ben01b}). These features have been explained by the
presence of a thick expanding shell above the photosphere. 

The present paper contains an extensive study of \g, an object which
sharing properties both of \z\/ and SNe IId reinforces the link
between these apparently different objects. It is worth to remind that 
a few other objects sharing some of the features of \g\/ have been
reported in recent years, i.e. SN~1997ab (Salamanca et al. 1998),
SN~1997eg (Salamanca et al. 2001) and SN~1998S (Fassia et al. 2000, 2001).

\begin{figure}

\psfig{file=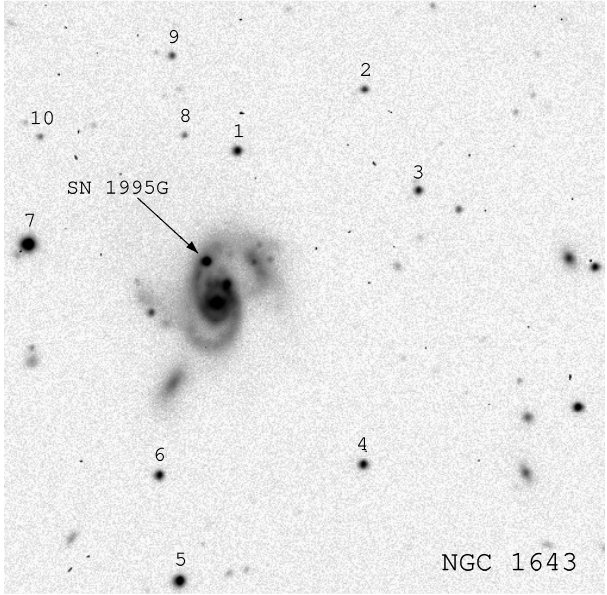,width=8cm} \caption{Identification of
SN 1995G and the stars of the local sequence around NGC1643 (see
Tab.\protect\ref{zzz}). North is up, East is to the left.}
\label{paperoga6}
\end{figure}

SN 1995G was discovered by Evans et al. \shortcite{evan95} on
Feb. 23.5 U.T., using the 1-m reflector of the Australian National
University in Siding Spring at an apparent visual magnitude of
15.5. McNaught \& Cass \shortcite{cass95} measured the SN position as
R.A. = $04^h43^m44^s.22$, Dec = $-05^{o}18'53''.8$ (equinox 2000.0),
4''.5 E and 16''.1 N from the center of NGC 1643 (see
Fig. \ref{paperoga6}), a Sbc galaxy, with $v_{helio}$= 4850$\pm$29
km/s \cite{huch93}.  Lacking other determinations, in the following we
will adopt the galaxy distance modulus derived from the recession
velocity, after correction for the Local Group infall into the Virgo
Cluster (LEDA\footnote{http://leda.univ-lyon1.fr}).  Assuming
$H_{0}$=65 km s$^{-1}$ Mpc$^{-1}$, we obtain $\mu$=34.32
(cfr. Sect.~4.2).

The basic information about \g\/ and the parent galaxy is summarized
in Tab. \ref{moe}.  It is worth mentioning that another type II
supernova, SN~1999et (Cappellaro, 1999b), has been discovered in the same
host galaxy.

We checked that no $\gamma$--ray burst has been detected by BATSE
in the months before the discovery within 2 error radii from the
position of \g. The closest event is burst 4B 950206B, which occurred
18 days before \g \/discovery \/at 2.33 BATSE error radii.

The observations of SN 1995G, described in Sect.~2, are analyzed in
Sect.~3 and 4. The discussion of the data is given in Sect.~5, and the
conclusions are summarized in Sect.~6.

\begin{table}
\caption{Main data on SN 1995G and the host galaxy.} \label{moe}
\begin{tabular}{|c|c|c|}
\multicolumn{3}{|c|}{SN 1995G} \\ \hline
$\alpha$ (J2000.0) & 04h43m44\fs22 & $\diamond$ \\
$\delta$ (J2000.0) & $-05$\degr18\arcmin53\farcs8 & $\diamond$ \\
Offset SN - Gal. Nucleus & 4''.5E, 16''.1N & $\diamond$ \\
SN Type & IIn & $\otimes$ \\
Discovery Date & 1995 Feb 23.5 & $\odot$ \\
(Julian Date) & (2449772) & $\odot$ \\
Discovery Magnitude & 15.5 & $\odot$ \\
&&\\
\multicolumn{3}{|c|}{NGC 1643} \\ \hline
$\alpha$ (J2000.0) & 04h43m44s & $\star$\\
$\delta$ (J2000.0) & $-05$\degr19\arcmin10\arcsec & $\star$\\
Morph. Type & SB(r)bc pec? & $\star$\\
Magnitude & 14.00 & $\triangle$ \\
Galactic Extinction $A_{B}$ & 0.19 & $\star$\\
Diameters & 1\farcm0 $\times$ 1\farcm0 & $\triangle$ \\
$v_{helio}$ (km s$^{-1}$) & 4850$\pm$29 & $\triangle$ \\
$\mu$ ($H_{0}$=65 km s$^{-1}$ Mpc$^{-1}$) & 34.32 & $\dag$\\
\hline
\end{tabular}

$\diamond$ \protect\cite{cass95}\\
$\otimes$ \protect\cite{fili95}\\
$\star$ \protect\cite{deva91}\\
$\odot$ \protect\cite{evan95}\\
$\triangle$ \protect\cite{huch93}\\
$\dag$ LEDA\\

\end{table}

\begin{figure}
\psfig{file=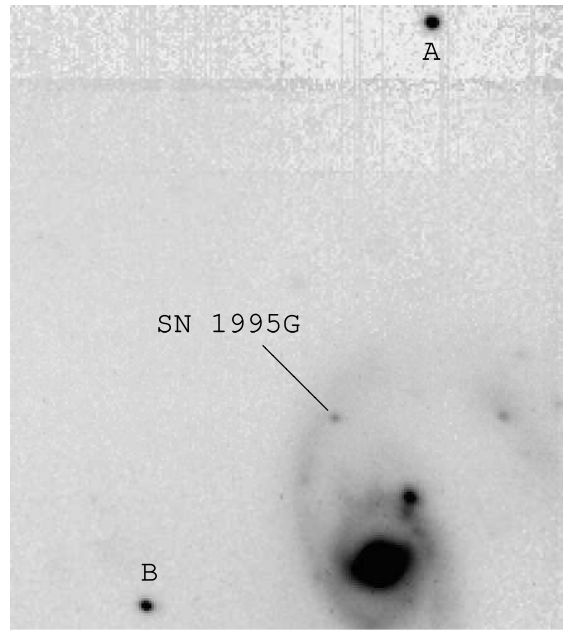,width=7.6cm}
\caption{SN 1995G and the IR local sequence (J band image, taken with ESO
2.2m + IRAC2 on 1996 November 30).  Note that star A coincides with
star 1 of the optical sequence. IR magnitudes of the stars are given
in Tab.\protect\ref{azz}.}
\label{willie}
\end{figure}

\section{Observations}
The photometric observations of \g\/ in the optical bands cover a
period of about 2.5 years.  They were carried out using different ESO
telescopes at La Silla and the Asiago 1.82m reflector.

CCD frames were reduced in the IRAF environment applying bias and
flat-field corrections. At early phases, when the SN was bright, the
SN magnitudes were measured with a PSF-fitting technique, which was
preferred over aperture photometry because of the SN location inside a
spiral arm.  The late time photometry was obtained by means of a
template subtraction technique, i.e. subtracting from the frame to be
measured a late (5 years after discovery) image of the field after proper
geometric, photometric and PSF matching.

The SN photometry has been performed relative to a local sequence
(Fig. \ref{paperoga6}) calibrated during several photometric nights by
comparison with standard stars from the list of Landolt
\shortcite{Land92}.  The magnitudes of the local sequence and those of
the SN are reported in Tab. \ref{zzz} and Tab. \ref{speedy},
respectively. At the last epochs since the SN was not detected,
only upper limits are reported.

Errors on the magnitudes of the local sequence stars are estimated
as the r.m.s. of the available measurements (no error is reported if
only one measure is available). Photometric errors of the SN were
estimated for some representative epochs (col.~6 of Tab. \ref{speedy}). 
They have been obtained via artificial star experiments, placing stars 
of the same magnitude as that of the SN in locations close to that of 
the SN, then computing the deviations of the recovered magnitudes.

\begin{table*}
\footnotesize
\caption{Magnitudes of the local sequence as identified in
Fig.\protect\ref{paperoga6}.} \label{zzz}  
\begin{tabular}{|c||c|c|c|c|c|}
 & 1 & 2 & 3 & 4 & 5 \\ \hline
U & 18.26$\pm$0.09 & 18.74 & 18.83 $\pm$0.04 & 18.53 $\pm$0.13 &
17.37$\pm$0.08 \\
B & 18.11$\pm$0.02 & 18.95$\pm$0.02 & 18.65$\pm$0.02 & 18.19$\pm$0.02 &
17.08$\pm$0.02 \\
V & 17.36$\pm$0.01 & 18.36$\pm$0.02 & 17.94$\pm$0.02 & 17.42$\pm$0.01 &
16.31$\pm$0.01 \\
R & 16.89$\pm$0.01 & 18.01$\pm$0.01 & 17.46$\pm$0.02 & 16.95$\pm$0.01 &
15.81$\pm$0.01 \\
I & 16.47$\pm$0.01 & 17.68$\pm$0.01 & 17.08$\pm$0.02 & 16.52$\pm$0.01 &
15.38$\pm$0.03 \\
\end{tabular}
\begin{tabular}{|c||c|c|c|c|c|}
 & 6 & 7 & 8 & 9 & 10 \\ \hline
U & 17.97$\pm$0.03 & 16.21 & 19.54 & 20.00 & 20.03 \\
B & 17.99$\pm$0.01 & 15.59$\pm$0.01 & 19.69$\pm$0.02 & 19.62$\pm$0.06 &
21.30$\pm$0.10 \\
V & 17.42$\pm$0.01 & 14.70$\pm$0.02 & 19.43$\pm$0.03 & 18.80$\pm$0.02 &
19.51$\pm$0.01 \\
R & 17.06$\pm$0.01 & 14.18$\pm$0.01 & 19.28$\pm$0.03 & 18.33$\pm$0.02 &
18.34$\pm$0.02 \\
I & 16.71$\pm$0.01 & 13.69$\pm$0.01 & - & - & - \\ \hline
\end{tabular}
\end{table*}

\begin{table*}
\caption{Optical and IR photometry of SN 1995G.} \label{speedy}
\small
\begin{tabular}{|c|c|c|c|c|c|c|c|c|c|c|c|} \hline
Date      & JD      & U     & B     & V     &$\Delta($V)&R   & I     &   J &  H & K' &Instrument \\ 
\hline\hline
28/02/95  & 49776.6 &  -    & 16.40 & 16.00 &   -     &15.73 & 15.63 &    -  &  -   &   -  & 0 \\
06/03/95  & 49782.6 & 15.96 & 16.33 & 16.02 &  0.03   &15.76 & 15.56 &    -  &  -   &   -  & 1 \\
30/03/95  & 49807.5 & -     & 16.58 & 16.26 &   -     &  -   &   -   &    -  &  -   &   -  & 2 \\
26/09/95  & 49986.6 & -     & 18.81 & 18.07 &   -     &17.41 & 16.95 &    -  &  -   &   -  & 0 \\
14/10/95  & 50004.8 & -     & 18.86 & 18.17 &   -     &17.50 & 17.21 &    -  &  -   &   -  & 2 \\
24/10/95  & 50014.6 & -     & 18.91 & 18.23 &   -     &17.59 &   -   &    -  &  -   &   -  & 0 \\
24/10/95  & 50015.5 & -     & 18.88 & 18.19 &   -     &17.56 &   -   &    -  &  -   &   -  & 0 \\
14/11/95  & 50035.5 & -     & -     & 18.43 &  0.05   &17.83 & 17.41 &    -  &  -   &   -  & 3 \\
22/11/95  & 50044.4 & -     & 19.10 & 18.34 &   -     &17.85 &   -   &    -  &  -   &   -  & 0 \\
24/11/95  & 50045.5 & -     & 19.03 & 18.47 &   -     &17.87 &   -   &    -  &  -   &   -  & 0 \\
17/01/96  & 50100.3 & -     & -     & 18.66 &   -     &18.07 &   -   &    -  &  -   &   -  & 0 \\
19/01/96  & 50101.6 & -     & 19.25 & 18.66 &   -     &18.11 & 17.77 &    -  &  -   &   -  & 4 \\
18/02/96  & 50131.6 & -     & 19.24 & 18.74 &   -     &  -   &   -   &    -  &  -   &   -  & 4 \\
22/02/96  & 50135.6 & -     & -     & -     &   -     &18.03 & 17.82 &    -  &  -   &   -  & 1 \\
04/09/96  & 50330.9 & -     & 19.83 & 19.38 &  0.10   &18.67 &   -   &    -  &  -   &   -  & 4 \\
06/09/96  & 50332.8 & -     & -     & -     &   -     &18.64 & 18.53 &    -  &  -   &   -  & 4 \\
19/11/96  & 50406.7 & 20.00 & 19.89 & 19.54 &   -     &18.83 & 18.62 &    -  &  -   &   -  & 1 \\
30/11/96  & 50418.3 & -     & -     & -     &    -    &  -   &   -   & 18.3  & 17.7 & 17.2 & a \\
02/01/97  & 50450.7 & -     & 19.93 & 19.65 &   -     &18.98 & 18.80 &    -  &  -   &   -  & 1 \\
24/01/97  & 50473.1 & -     & -     & -     &    -    &  -   &   -   & 18.5  & 18.2 & 17.6 & a \\
28/03/97  & 50534.1 & -     & -     & -     &    -    &  -   &   -   & 18.4  & 18.4 & 17.4 & a \\
17/09/97  & 50709.4 & -     & -     & -     &    -    &  -   &   -   & 19.1: &  -   &   -  & a \\
22/09/97  & 50713.9 & -     & 21.27 & 21.44 &  0.40   &20.22 & 20.14 &    -  &  -   &   -  & 4 \\
17/11/97  & 50770.7 & -     & -     & -     &    -    &  -   &   -   & 19.2  &  -   &   -  & a \\
22/12/98  & 51169.8 & -     & -     & $\ge$21.4& -    &$\ge$21.2 & - &    -  &  -   &   -  & 1 \\
03/01/99  & 51182.7 & -     & -     & -     &    -    &  -   &   -   & 20.7  &  -   &   -  & b \\
04/11/99  & 51486.8 & -     & -     & -     &    -    &$\ge$22.1 & - &    -  &  -   &   -  & 5 \\ 
\hline
\end{tabular}

0 = Asiago 1.82m; 1 = Dutch 0.91m; 2 = ESO 3.6m + EFOSC1; 3 = NTT 3.5m
+ EMMI; 4 = ESO 2.2m + EFOSC2;\\ 
5 = Danish 1.54m + DFOSC; a = ESO 2.2m + IRAC2; b = NTT + SOFI
\end{table*}


In addition to optical data, a number of late infrared observations
have been obtained. For each epoch, a sequence of SN and nearby sky
frames are available. After flat--field correction and sky
subtraction, the frames were aligned and combined. For the IR photometric
calibration we used standard stars taken from Carter \& Meadows
\shortcite{cart95}, Hunt et al. \shortcite{hunt98} and Persson et al.
\shortcite{pers98}.
The local sequence consists of only two stars in the field (see
Fig. \ref{willie} and Tab. \ref{azz}). The IR magnitudes of \g\/ are
reported in Tab. \ref{speedy}. Typical errors of the SN infrared magnitudes 
 are 0.2 mag (0.4 mag on Sept. 17, 1997).

\begin{table}
\caption{IR magnitudes for stars of the local sequence.}
\label{azz}
\begin{tabular}{|c||c|c|} \hline
Filter & A & B \\ \hline\hline
J & 15.88$\pm$0.07 & 16.32$\pm$0.02 \\
H & 15.40 & 15.60$\pm$0.02  \\
K' & - & 15.52$\pm$0.05 \\ \hline
\end{tabular}
\end{table}

%

Optical spectroscopic observations were obtained at La Silla, Asiago
and Lick Observatories. The log of these observations is given in
Tab. \ref{burt}.  Wavelength calibrations were performed using
comparison spectra of HeAr lamps while the flux calibration was done
using spectra of standard stars (Stone \& Baldwin, 1983; Baldwin \&
Stone, 1984; Hamuy et al., 1992; Hamuy et al., 1994) taken during the
same nights.  We made also an attempt to correct the SN spectra for
atmospheric absorptions lines by comparison with the observed spectrum
of a featureless hot star.  Finally, the flux calibration of the
reduced SN spectra was checked against the broad band photometry and
found in reasonable agreement (typical error is $\pm$20-30\%).

\begin{figure*}
\psfig{file=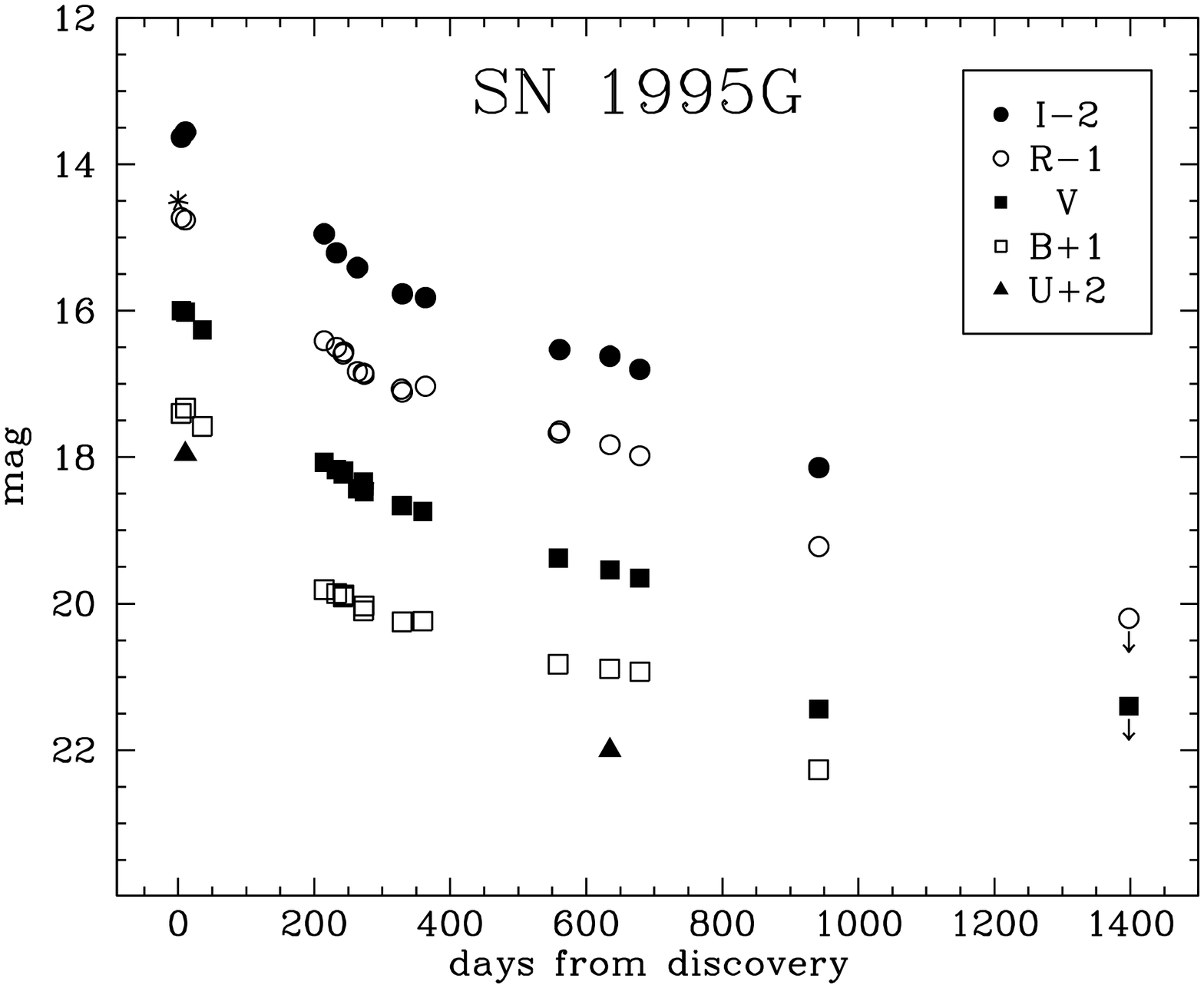,width=16cm,angle=270} \caption{Light
curves of SN 1995G in B, V, R and I bands. The two U points are also
reported. The asterisk refers to the discovery unfiltered magnitude
(IAUC 6138) plotted on the same scale as R photometry. } \label{lc}
\end{figure*}

\section{Photometry}\label{phot}
\subsection{Optical Light and Colour Curves}\label{lcsect}

The BVRI light curves of \g\/ (plus 2 measurements in U) are shown in
Fig. \ref{lc}. Though there are not definite evidences, we believe
that the SN was discovered very close to maximum brightness. The blue
color of the first spectrum is consistent with this belief
(cf. Sect.~4).

\begin{table}
\caption{Decline rates (mag/$100d$) of SN 1995G.}
\label{autogatto}
\footnotesize
\begin{tabular}{|c|c|c|c|c|}\hline
$\Delta$(Days) & $\gamma_{B}$ & $\gamma_{V}$ & $\gamma_{R}$ & $\gamma_{I}$
\\ \hline \hline
0-215   & 1.19 & 1.07 & 0.80 & 0.65 \\
215-680 & 0.25 & 0.33 & 0.31 & 0.36 \\
680-940 & 0.51 & 0.68 & 0.47 & 0.51 \\
\hline
\end{tabular}
\end{table}

At all wavelengths the light curves are relatively flat, decreasing by
only 5 magnitudes in 2.5 years. The slopes of the light curves
computed in different time intervals are given in
Tab. \ref{autogatto}. During the first months the slopes are steeper
and then level out with time, until 1--2 years after the discovery
they are as flat as 0.2-0.3 mag/$100d$ in all optical bands. A
steepening in the luminosity decline was observed only after 650 days
(about 0.5 mag/$100d$).  A part from that, as we will see later, the
light curve is very similar to that of \z\/ \cite{tura93}.

\begin{figure}
\psfig{file=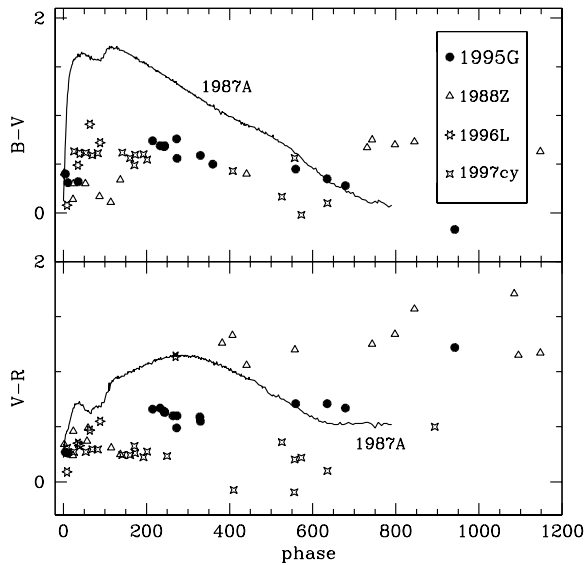,width=8.5cm,angle=0}
\caption{Top: $(B-V)$ color evolution for SN 1995G compared with SN 1988Z,
SN 1996L, SN 1997cy and SN 1987A; bottom: $(V-R)$ color evolution for
the same objects.}  \label{color}
\end{figure}

Also, the colour evolution of \g\/ (Fig.\ref{color}) is similar to
that of \z. Because of the cooling of the expanding layers,
most ``normal'' (non--interacting) SNe II show a rapid reddening after
the explosion and only the emergence of nebular lines makes the colour
turn blue again \cite{pata94}.  In \g\/ instead we observe a slow
increase in $(B-V)$ from 0.3 to 0.7 in $\sim 250$ days, then the colour
curve flattens somewhat. Allowing for the large uncertainties, the SN
seems to become blue again at about 2.5 years after the discovery. It
is interesting to note that the color curves of other SNe IIn, SN
1988Z (Turatto et al., 1993; Aretxaga et al., 1999) and SN 1997cy
(Turatto et al., 2000), show similar properties, while the few points
of SN 1996L (Benetti et al., 1999) indicate a more rapid evolution.

\subsection{The Absolute and Bolometric Light Curves}

\begin{figure}
\psfig{file=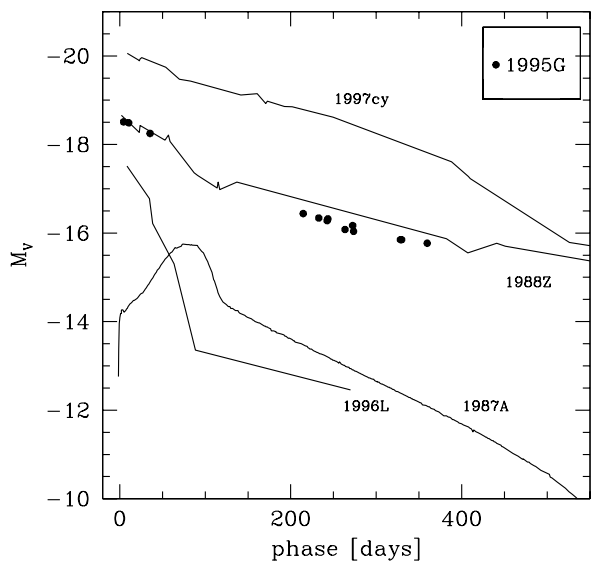,width=8.5cm,angle=0}
\caption{Absolute V light curve of SN 1995G compared with that of
SN 1988Z (IIn, Turatto et al. 1993), SN 1997cy (IIn, Turatto et al. 2000),
SN 1987A (II, Patat et al. 1994, and
references therein) and SN 1996L 
(IId, Benetti et al. 1999).} \label{abs}
\end{figure}

Adopting the distance modulus $\mu$=34.32 and a total absorption
$A_{B}$=0.20 (cfr. Sect.~\ref{radvel}), we can calculate the absolute
magnitudes near maximum. The values are: $M_{B}\sim-18.19$, $M_{V}\sim
-18.47$, $M_{R}\sim-18.72$ and $M_{I}\sim-18.85$.  Figure~\ref{abs}
shows the comparison of the absolute V light curve of \g\/ with those of other
representative SNe II. It is remarkable the close coincidence with \z.

In order to compute the energy budget of SN 1995G and to simplify the
comparison with theory, it is useful to determine the bolometric light
curve.  Unfortunately no detections at UV, X--ray or radio wavelengths
are available and we stress that in the case of \z\/ the contributions
at these wavelengths were important \cite{aret99}. Actually, the
position of \g\/ was targeted with the VLA on 1995 June 22 (Van Dyk,
private communication), but no radio emission was detected.
 
The available data range from U to K' bands with much denser sampling
at optical wavelengths.  U data are available only at two distant
epochs. We have estimated the contribution at intermediate epochs
interpolating the U band contribution to the total emission between
the two available epochs. Unfortunately near--IR observations start
only when those in the U band cease, that is after 600 days.

\begin{figure}
\psfig{file=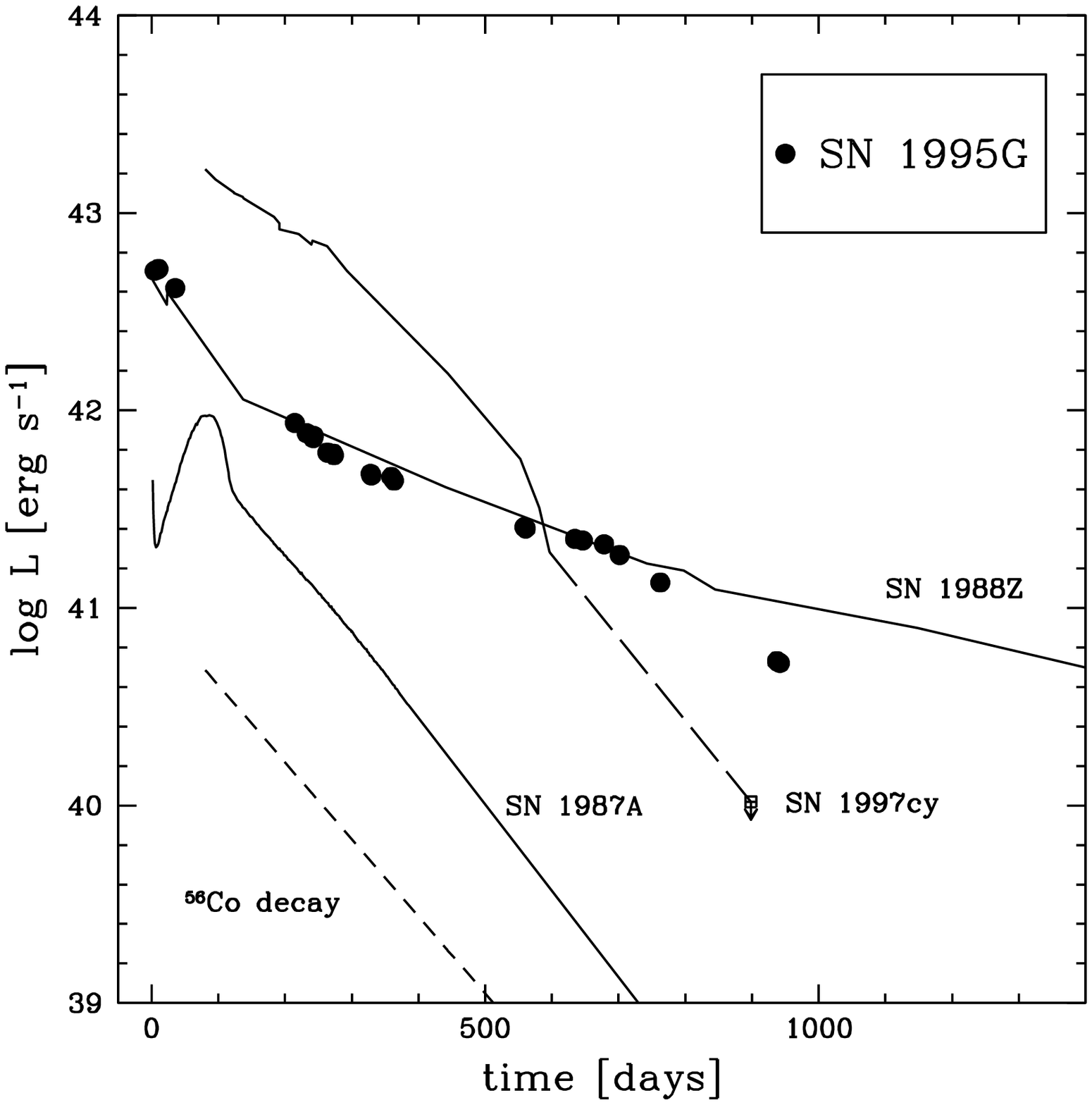,width=8.5cm,angle=0}
\caption{The
UBVRI bolometric light curve of SN 1995G (dots) compared to
that of SN 1987A (UBVRIJHKL, Catchpole et al., 1987, 1988, 1989), SN
1997cy (UBVRI, Turatto et al. 2000) and SN 1988Z (BVR, Aretxaga
et al. 1999). The decline rate of $^{56}$Co is reported for
reference. The dashed tail of SN 1997cy connects the last detection to
an upper limit and therefore the true decline can be steeper than that
drawn here.}
\label{homer}
\end{figure}

The $UBVRI$ bolometric light curve is shown in Fig.~\ref{homer} along
with those of other SNe IIn and that of SN 1987A.  As a reference we
note that the luminosity of SN 1995G at the time of the first
observation is only 5 times that of SN 1987A at the secondary maximum,
while after two years it is 200 times brighter.  Note that these
latter are lower limits to the ratio of the bolometric luminosities
because in the case of SN 1987A the IR contribution is included. The
$UBVRI$ curves of SN 1995G and SN 1988Z \cite{aret99} are very
similar. SN~1997cy, to date the most luminous type II SN, had a much
faster decline rate and indeed after day 600 \g\/ outshines SN~1997cy.

Since IR observations of \g\/ are available only at
late epochs, their contribution to the energy budget immediately
after the SN explosion is unknown. We note, however, that around day
640 after the discovery, when we have the broadest photometric
coverage from U to K', the cumulative contribution of J, H, K' bands
is about 50$\%$ that of UBVRI. Our data indicate that the
contribution of the near--IR increases with time, because the decline rates 
of infrared light curves appears to be slower than the optical ones.

We note that our IR observations of \g\/ in the interval between day
640 and day 750 do not show the rise of the K' continuum detected by
Fassia et al. \shortcite{fass00} for the type IIn SN~1998S and
attributed to emission from pre-existing, warm dust in the
circumstellar material.


\section{Spectroscopy}

\begin{table*}
\caption{Spectroscopic observations of SN 1995G.} \label{burt}
\begin{tabular}{cccccccc}\hline
Date & JD & Days & Instrument & Grism or & Resol. & Exp. & Range \\
 & & from disc. & & grating & ( A) & (min.) & ( A) \\ \hline
25/02/95 & 49773.5 & 2  & Lick       & 600,1200,600 & 7,4,7 & 7.5,7.5,7.5
& 3380-7300 \\
30/03/95 & 49807.5 & 36 & ESO3.6+EF.1 & B300,O150,R300 & 18.4,7.6,21.5 &
10,15,10 & 3715-9805 \\
14/10/95 & 50004.9 & 233 & ESO3.6+EF.1 & B300 & 18.4 & 15 & 3745-6940 \\
31/10/95 & 50021.6 & 250 & As1.82+B$\&$C & 150 & 23 & 60 & 4530-8305 \\
14/11/95 & 50035.7 & 264 & NTT+EMMI & gm$\#$3 & 7.8 & 30 & 3860-8390 \\
19/01/96 & 50101.7 & 330 & ESO2.2+EF.2 & gm$\#$1,gm$\#$6 & 36,11.2 & 30,30
& 3325-9215 \\
18/02/96 & 50131.6 & 360 & ESO2.2+EF.2 & gm$\#$6 & 11.2 & 60 & 4450-7060
\\
06/09/96 & 50332.5 & 561 & ESO2.2+EF.2 & gm$\#$3,gm$\#$5 & 8.8,10.8 &
30,30 & 3220-9250 \\
19/12/96 & 50436.5 & 665 & ESO1.5+B$\&$C & gt$\#25$ & 8 & 60 & 3515-9280
\\
17/02/97 & 50496.6 & 725 & ESO1.5+B$\&$C & gt$\#15$ & 10 & 60 & 3085-10640
\\
22/09/97 & 50713.9 & 942 & ESO2.2+EF.2 & gm$\#$5 & 11 & 60 & 5205-9305 \\
\hline
\end{tabular}
\end{table*}

\subsection{The Spectral Evolution}

\begin{figure*}
\psfig{file=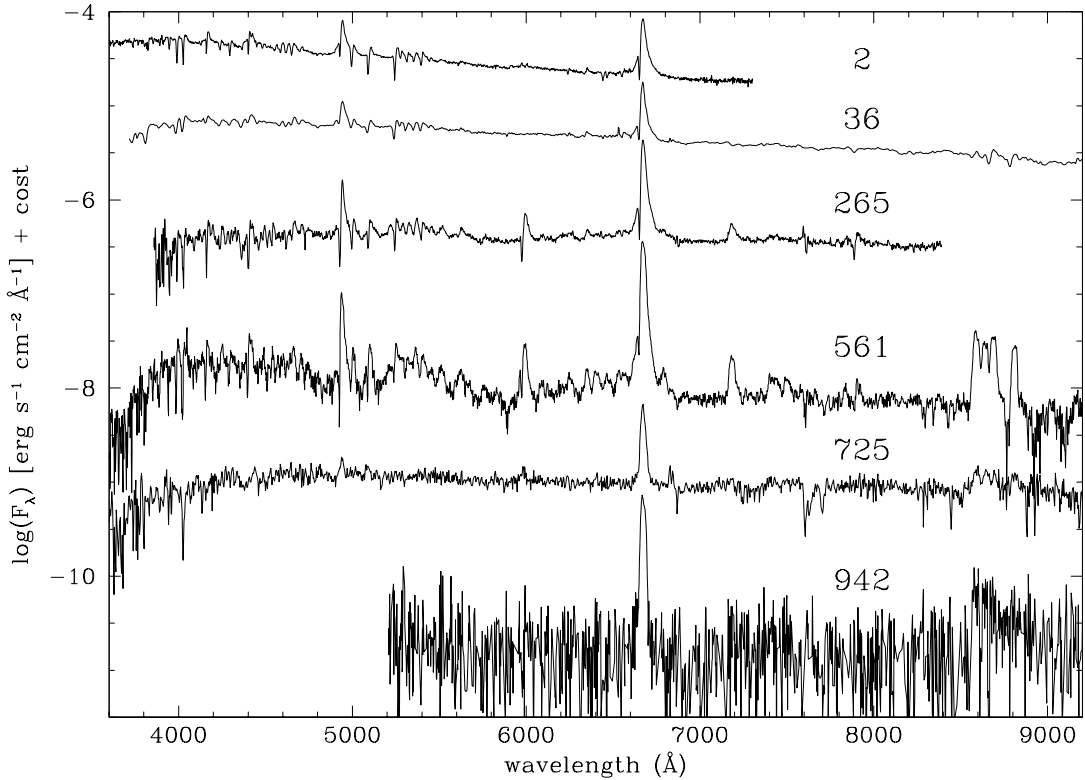,width=16cm,angle=270} \caption{Spectral
evolution of SN 1995G. The spectra are in the observer rest frame and
the days from discovery are reported near each spectrum.
On day 725 the atmospheric absorptions could not be removed.
In combining spectra, priority has been
given to those of higher resolution.} \label{pluto1}
\end{figure*}

The spectral evolution of \g\/ during the first 2.5 years after the
explosion is shown in Fig.~\ref{pluto1}. The first spectrum was taken
within two days from the discovery \cite{fili95}. It shows a
relatively blue continuum (T$_{eff}$=8800 K after de--reddening,
cfr. Sect.~\ref{radvel}) with relatively narrow Balmer P--Cygni
features superimposed on broader emission components.  The overall
intensity ratio of \Ha/\Hb\/ is 1.58, lower than Case B recombination
(2.8) while the intensity ratios of the individual components
(cfr. Sect.~4.4) are very difficult to measure.  The departure from
simple recombination is the result of line formation in an extended
atmosphere where the underlying continuum flux is stronger at \Hb\/
than at \Ha.  The P--Cygni absorption minima are displaced by about
700 km s$^{-1}$ from the emission-line cores both for H and Fe~II
lines.

The next spectrum, taken about one month later, is similar, with
a continuum temperature T$_{eff}$=7000 K and several narrow P-Cyg
lines of H, O~I, Ca~II and Fe~II (see Sect.~4.3).  Neither He nor
Na~I~D lines are visible. The most evident difference with respect to
the first spectrum is in the \Ha/\Hb\/ ratio which has now increased
to 3.0.  Again this can be understood to result from the decrease of
the continuum flux at \Hb\/ relative to \Ha.  Indeed, in the same
period the ratio of the equivalent widths does not change
significantly (EW(H$\alpha$)/EW(H$\beta$)=3.0 and 3.2, respectively).
Though conspicuous, the emission lines in the early spectra account
for only about 4\% of the total optical flux and have only a small
influence on the SED.

The SN was then recovered eight months after discovery when the
spectrum had noticeably changed. The continuum was redder and fainter,
and dominated by emission lines of H$\alpha$, H$\beta$ and the
infrared Ca II triplet. He I lines (e.g. 7065 A) appear and remain
clearly visible for about 1 year, but contrary to other lines they do
not show a blueshifted absorption. The He I 5876 A line is blended
with Na~I~D, the latter having the narrow P--Cygni component. The very
late spectrum shows only Balmer lines in emission (and probably IR Ca
II). It should be noted that the galaxy contribution to the spectrum
is not completely removed when the SN fades and the spectra are
contaminated by the unresolved [S II] 6717,6731 A doublet, coming from
an underlying H II region. Probably the background contaminates also
the H$\alpha$ narrow component with a (small) contribution to the
total flux of the line. It is remarkable that during all the evolution
there is a monotonic increase of the Balmer decrement.

\begin{figure*}
\psfig{file=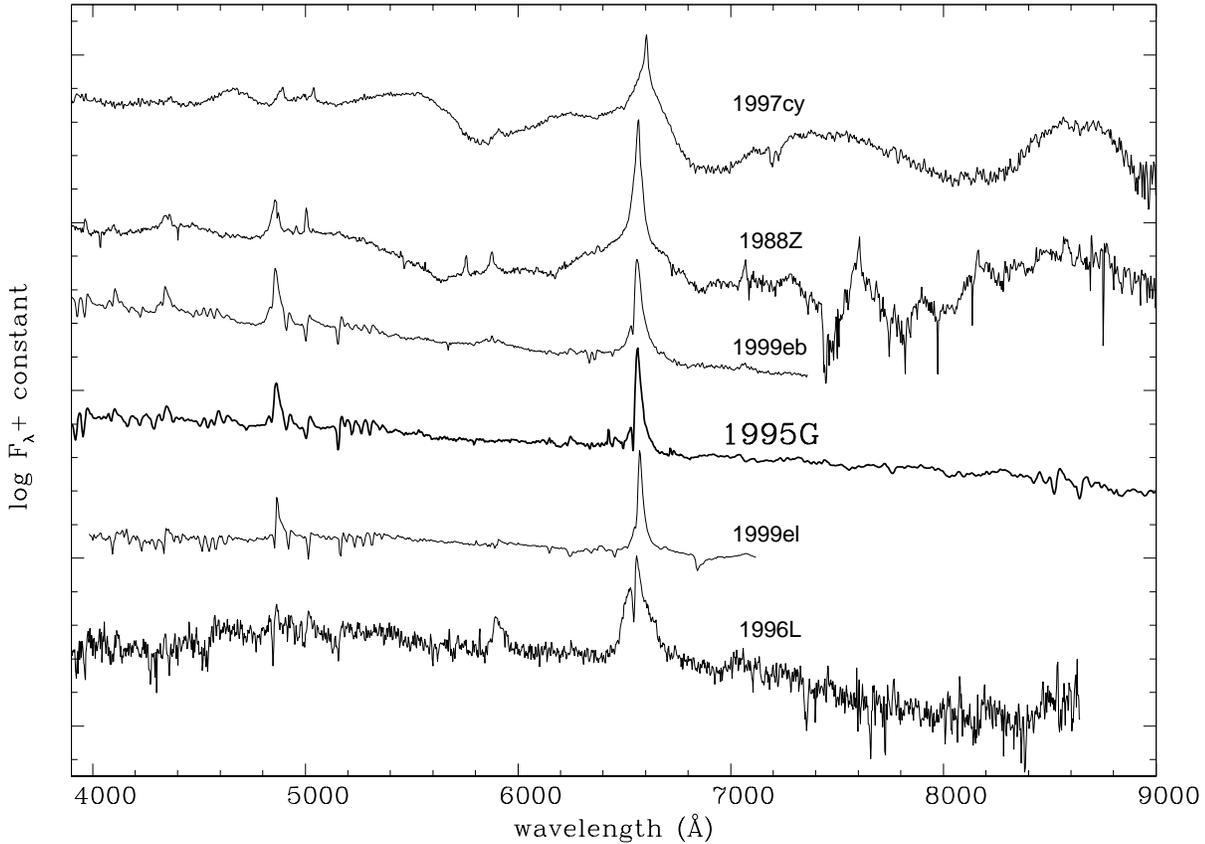,width=17.5cm,angle=270}
\caption{Comparison among the spectrum of SN 1995G taken on 1995
March 30 (36 days after the discovery) and spectra of SN 1997cy (39 days), 
SN 1988Z (4 months), SN 1999eb (2 months), SN 1999el (18 days), 
SN 1996L (2 months) and SN 1994aj (50 days). All spectra are 
in the parent galaxy rest frames.}
\label{cfr_sp}
\end{figure*}

In Fig. \ref{cfr_sp} we compare the spectrum of \g\/ taken on 1995
March 30 with those of other SNe IIn.  Despite the general similarity,
the spectrum shows important differences with both \cy, and \z.  The
line widths of these two objects are much broader than those of \g,
implying expansion velocities of about 15000 \kms\/ instead of
4-5000\kms\/, hence probably higher explosion energies (Turatto et
al., 2000; Aretxaga et al., 1999). The fact that \g\/ and \z\/ have
different kinematics of the ejecta, but nearly identical broad band
light curves and overall luminosity evolution (Fig.~\ref{homer}) must
be connected to the different densities of the respective
circumstellar media (cfr. Sect.~5).

The spectrum of SN 1995G more closely resembles that of SN 1999eb taken 2 
months after the discovery, and SN 1999el taken 18 days after the discovery 
(see Fig.~\ref{cfr_sp}). They have similar continua and strong emission lines 
with similar profiles. Curiously, the light curves of these objects
having strong spectral similarities to \g\/,
are steeper than that of \g, as shown by sparse unpublished observations from 
our archive. 
 
In recent years several SNe have shown narrow P-Cygni absorptions.  As
we mentioned before, a number of SNe, sometimes dubbed IId or IIsw,
show these features over an otherwise normal, broad--line photospheric
spectrum. The spectrum of one of these, SN~1996L \cite{bene99} is
shown in Fig.~\ref{cfr_sp}.  It has been argued that in all these
cases the narrow P-Cygni features originate in a slowly expanding
shell of gas which has been ejected shortly before the explosions.  In
spite of the fact that these SNe show spectroscopic signature of
eject-CSM interaction, they have light curves that decline as steeply
as typical linear SNe II, and hence differ from
\g.

\subsection{The Recession Velocity and the Interstellar Absorption}\label{radvel}

The recession velocity of the host galaxy at the location of the SN,
$v_{h}=4911\pm70$ \kms, has been derived from the narrow emission
lines (\Ha, \Hb, [N~II] and [O~III]) arising from the H II
regions surrounding the SN and measured on the spectrum of better
resolution.  This value is consistent with the nuclear velocity of NGC
1643, $v_{h}=4850\pm29$ km s$^{-1}$ \cite{huch93}, since SN 1995G is
located in an outer arm of the parent galaxy.

Interstellar absorption lines of Na~I~D are not clearly detected
neither at the rest wavelength nor in the parent galaxy rest--frame,
suggesting that the reddening cannot be particularly strong. The values
of Galactic absorption by Schlegel et al. \shortcite{schl98} and RC3
are in good agreement, respectively $A_{B}^{g}=0.22$ and 0.19. In the
following we will adopt for \g\/ a total extinction $A_{B}=0.20$,
assuming therefore no absorption within the parent galaxy.

\subsection{Line Identification}
\begin{figure*}
\hspace{3cm} \psfig{file=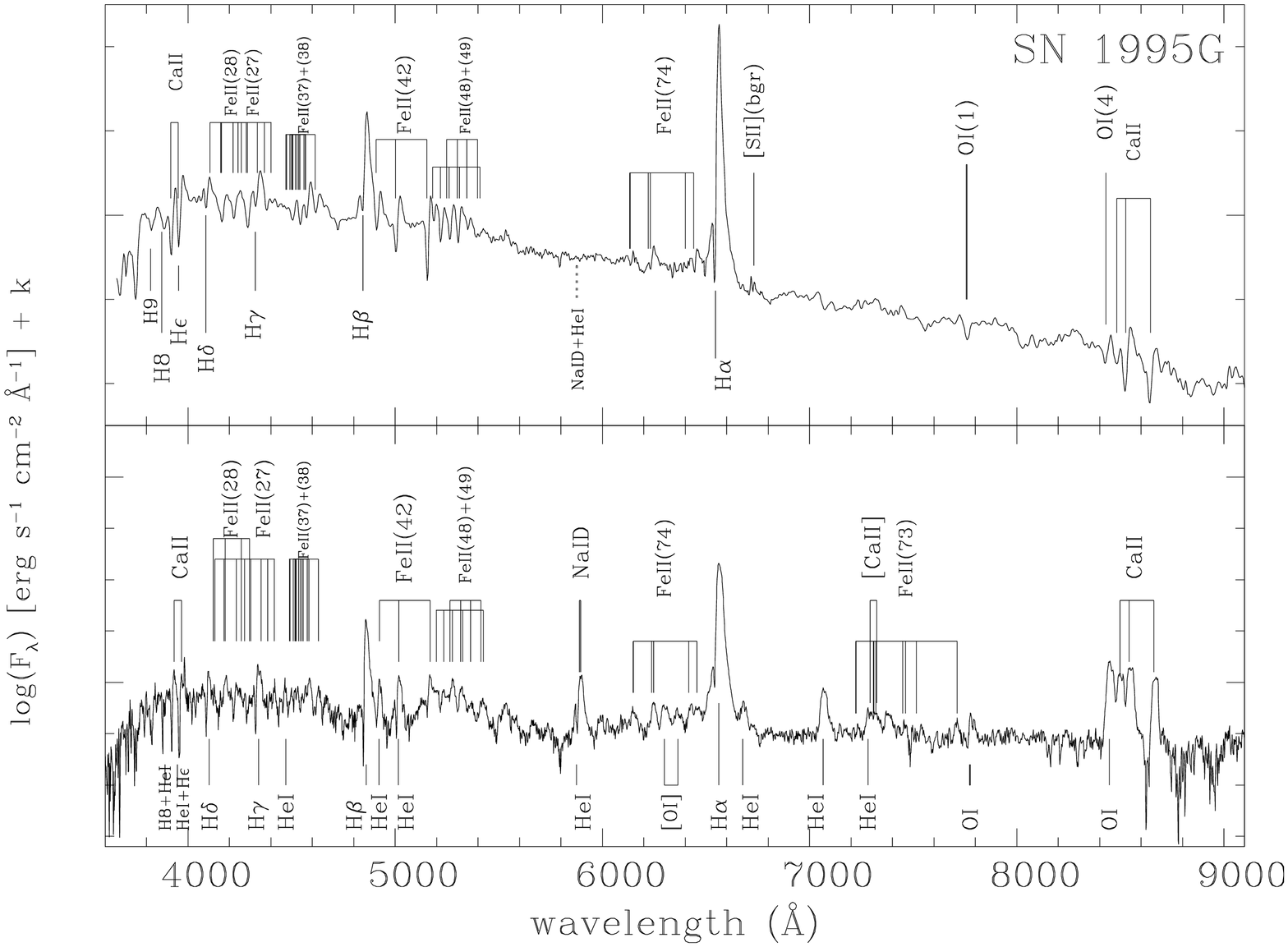,width=15.0cm,angle=270}
\caption{Line identifications in the spectra of SN~1995G. Top:
early--time spectrum taken on 1995 March 30. The vertical lines
mark the minima of P-Cygni absorptions, blueshifted by 750 \kms\/ with
respect to the line rest wavelengths. Bottom: late spectrum on
1996 September 6. Marked are the cores of the emissions. The
spectra are blueshifted to the host galaxy rest frame.}
\label{lupin}
\end{figure*}

The line identifications have been made on two good
signal--to--noise spectra taken on day 36 (1995 March 30) and day 561
(1996 September 6). The identifications are shown in Fig.~\ref{lupin}.
 
At early times the strongest features are the H~I Balmer lines
characterized by broad emissions on top of which are narrow P-Cygni
features.  Fe~II multiplet lines are also identified as absorption
features in the bluest part of the spectrum (cfr. Fig.~\ref{lupin}).
Ca~II H \& K, and its IR triplet are also present and resolved in the
multiplet components, as well as the O I lines at 7772 A\/ and 8446
A\/. Two emissions at about 6250 and 5530 A (not marked in the figure)
might be attributed to Sc~II. Noticeable is the absence at this epoch
of Na~I~D and He~I lines which will appear only later
(cf. Sect.~4.5). Unlike the case of \z\/ very high ionization
(coronal) lines are not detected.

Numerous lines are identified also at the late epoch. Because of the small
velocity of the emitting layers, the individual components of the IR Ca~II
triplet are resolved as well as the strong O~I 8446 A\/ emission. The strength
of this line and the uncertain presence of other oxygen lines (such as 
O~I 7772 A\/) is suggestive of the Ly$\beta$ pumping mechanism \cite{grandi}. 
[O~I] 6300, 6363 A\/ is not unambiguously identified because of the 
presence of other lines. Fe~II is still present in numerous multiplets, while 
the presence of [Fe~II] lines is not certain. In contrast with early spectra, 
many He~I emissions lines become visible; among these the most obvious 
are at 7065, 6678, 5876, 5015, 4471 A, none of which shows
the P--Cygni absorption component still visible in other lines.
The 5876 A\/ line is blended with Na~I~D.

\begin{figure*}
\psfig{file=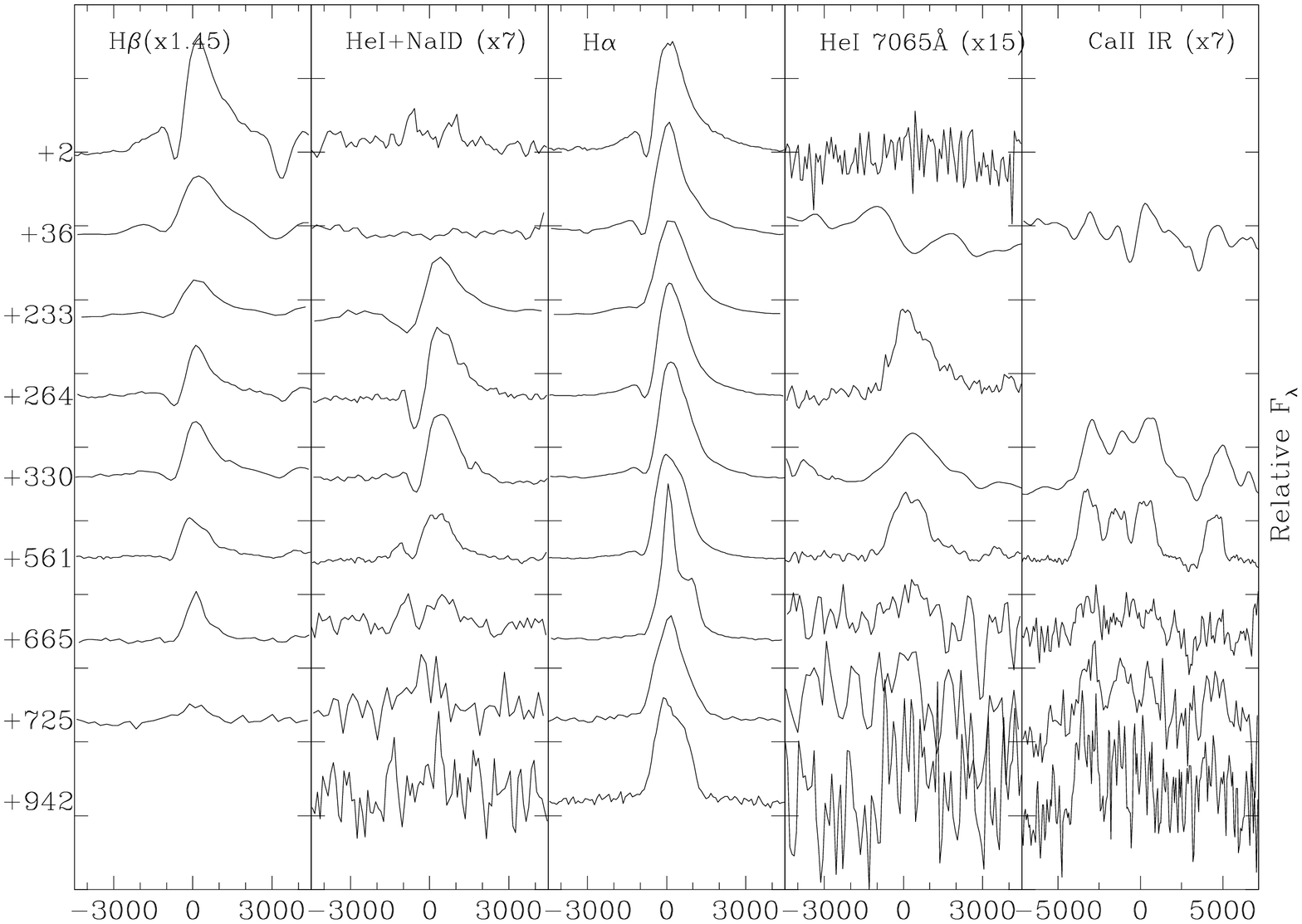,width=16cm,angle=270}
\caption{Evolution of the profiles of the main spectral lines of SN
1995G.
The peak on the red wing of \Ha\/ in the spectrum taken 665 days
after SN discovery is [N~II] due to the background. The
\Ha\/ lines are normalized at their peak; the other lines are normalized 
with \Ha\/ and then multiplied by a constant.} \label{lineprof}
\end{figure*}

\subsection{The Profile and Evolution of the Balmer Lines}\label{linprofsect}

\begin{table}
\caption{Expansion velocities from the minima of P-Cygni absorptions.}
\label{dylandog}
\begin{tabular}{cccc}
\hline
Date & days after & v(\Ha)& v(Fe~II 5169A) \\
     & discovery & (\kms) & (\kms)\\
\hline
25/02/95 & 2   & 710 & 510 \\
30/03/95 & 36  & 840 & 750 \\
14/10/95 & 233 & 750: & 600: \\
31/10/95 & 250 & -   & 630: \\
14/11/95 & 264 & 670 & 470 \\
19/01/96 & 330 & 620 & 470 \\
18/02/96 & 360 & 670 & 590 \\
06/09/96 & 561 & 750 & 800 \\
	\hline
\end{tabular}
\end{table}

\begin{table*}
\caption{Three--component Gaussian deblending of \Ha\/, and Balmer
decrements.} \label{gauss}
\begin{tabular}{cccccccccccccc}
\hline
Date & Phase & \multicolumn{4}{c}{FWHM$^*$(\Ha) [\kms]} &&
\multicolumn{5}{c}{flux$^\dagger$(\Ha) [$10^{-16}$ erg cm$^{-2}$ s$^{-1}$]}
& \multicolumn{2}{c}{Balmer decrement} \\
\cline{3-6} \cline{8-11}
  & [days] & broad & interm. & narrow & abs. & &broad & interm.& narrow &
  abs. & total & obs. & $\dagger$ \\
\hline
25/02/95 & 2 & 4150 & 1850 & 950 & 400 && 416& 530& 219 & -138& 1170 & 1.58
& 1.50 \\
30/03/95 & 36 & 4600 & 2100 & 700 & 400 && 226 & 373 & 225 & -75 & 824 &
3.03 & 2.89 \\
14/10/95 & 233 & & & & && & & & & 360 & 5.45 & 5.20 \\
31/10/95 & 250 & & & & && & & & & 524 & 5.35 & 5.09 \\
14/11/95 & 264 & 3100 & 1300 & 250 & 450 && 202 & 320 & 11 & -68 & 533 &
5.61 & 5.35 \\
19/01/96 & 330 & 3000 & 1200 & 100: & 450 && 197 & 277 & 25 & -59 & 729 &
5.52 & 5.26 \\
18/02/96 & 360 & 3450 & 1400 & unres. & 450 && 138 & 321 & 11 & -68 & 470 & 6.02
& 5.74 \\
06/09/96 & 561 & & & & & & && & & 531 & 6.32 & 6.02 \\
19/12/96 & 665 & 2200 & 1150 & unres. & 450: && 85 & 124 & 53 & -14: & 262 &
6.56 & 6.24 \\
17/02/97 & 725 & & & & & & && & & 176 & 12.57 & 11.97\\
22/09/97 & 942 & \multicolumn{3}{c}{Boxy(FWZI$\sim2500$)}& && & & &  & 116 & & \\
\hline
\end{tabular}
\begin{flushleft}
(*)deconvolved for the spectral resolution\\
$\dagger$ corrected for reddening \\
\end{flushleft}
\end{table*}

The line profiles of \Ha\/ and \Hb\/ are complex and evolving with
time. At early phases the lines show strong and narrow emission cores
with blue absorptions superposed on broad wings. The relative
intensities of the components change and the absorption features
completely disappear at about 2 years. The temporal evolution is
summarized in Fig.~\ref{lineprof} together with that of other
significant lines.

The expansion velocities at the photosphere derived from the P-Cygni
minima are reported in Tab.~\ref{dylandog}. The velocities measured for
\Ha\/ are of the order of 700--800 \kms\/ thus indicating that
either the kinetic energy of the ejecta is very small, as in the case of
SN~1997D (Turatto et al., 1998; Benetti et al., 2001) or that 
they arise from an outer, slowly expanding
shell as in the case of SNe 1994aj \cite{bene98} and 1996L
\cite{bene99}. The expansion velocities measured from the Fe~II 5169A\/
absorption are on average slightly lower, about 600 \kms. The typical
FW half depth of the \Ha\/ absorption lines is around 400-450 \kms
(cfr. Tab.~\ref{gauss}).  Contrary to the absorption lines of normal
SNe which form just above the receding photosphere (and hence show a
progressive red-shift of the minima), the absorptions of \g\/ remain at
constant location in wavelength.

To disentangle the different components of \Ha, we performed a
multicomponent Gaussian fit of the lines. Within the first two years
at least two emission components and a narrow blueshifted P-Cygni
absorption are required to get a reasonable fit. However, the fitting
of spectra of higher signal--to--noise and resolution requires a third
emission component. The fitting was performed using the deblending
option of the {\em splot} command in IRAF. A part for giving an inital
guess for the position, FWHM and peak intensity of each component, no
additional constraints were imposed.  The FWHM and fluxes of the
components of H$\alpha$ are shown in Tab.~\ref{gauss}.  Although we
are aware that fitting multiple Gaussian components to the composite
spectral profile is somewhat arbitrary, we believe that it is safe to
conclude that the observed emission lines arise from regions with
significantly different characteristic velocities.

From the Table we note a slow monotonic evolution of the FWHM(\Ha).
The broad component has a FWHM of 4000--4500 \kms\/ in the first
spectra, decreasing to about 3000 \kms\/ after 1 year and to 2200
\kms\/ after 2 years. The intermediate component also
declines from about 2000 to 1200 \kms. The narrow component is
resolved in the early spectra, while at later epochs the line width is
below the spectral resolution. The evolution in width and intensity of
this narrow component demonstrates that it is not due the galaxy
background although some contamination is possible (cfr.  the [N~II]
6584A emission on day 665, Fig.~\ref{lineprof}).  At the last epoch
(942 days) the Gaussian fit fails because the profile is clearly boxy
with FWZI$=2500$ \kms.

\begin{figure}
\psfig{file=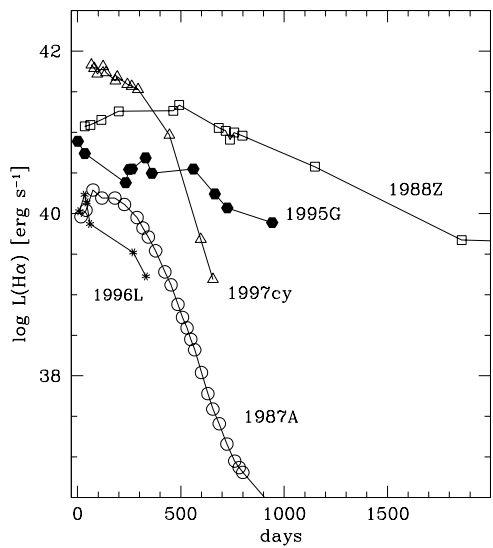,width=8.5cm,angle=0} \caption{The
evolution of the \Ha\/ luminosity for SN 1995G and other CSM
interacting SNe, \z\/, SN 1997cy and SN 1996L. Also SN 1987A is
reported for comparison.
Typical error in logL(\Ha) is 0.3.} \label{hal}
\end{figure}

The total \Ha\/ flux, as well as the fluxes of the individual
components, show an overall decline with time.  Small scale
fluctuations in the \Ha\/ flux are to be ascribed to uncertainties in
the absolute flux calibration and/or to uncertainties in the fitting.
The comparison of the \Ha\/ integrated luminosity evolution of \g\/
with that of other SNe (Fig. \ref{hal}) shows that \g\/ is similar to
the CSM--interacting type IIn rather than to the
radioactivity--powered normal SNe II, e.g. SN~1987A.  We note however,
that while the $UBVRI$ bolometric light curve of \g\/ was nearly
identical to that of \z\/ (Fig.~\ref{homer}), the \Ha\/ flux is about
0.8 dex fainter.

\subsection{He~I Lines}

He~I emission lines are not visible in the first spectra (cfr.
Fig.~\ref{pluto1} and \ref{lineprof}). They emerge only after day 200,
reach the maximum intensity between 8 and 20 months, and then
disappear. In the 561 day spectrum, when He emissions reach the
maximum relative intensity, we identify He~I 4471 A\/, 6678 A\/ and
7065 A\/. None of these lines show evidence of an absorption
component. Other He~I lines are probably blended with lines of other
ions, e.g. with Fe~II 4921 A\/ and 5015 A\/. In particular, He~I 5876 A\/
is blended with the Na~I~D P-Cygni feature.

For the clean 7065 A\/ emission we measure a FWHM velocity between
1200-1800 km s$^{-1}$, close to that of the intermediate component of
the Balmer lines.  The flux of this line is about $3 \times 10^{-15}$
erg cm$^{-2}$ s$^{-1}$, between day 250 and 561. In the spectrum of
day 665 the flux is one order of magnitude smaller (about $3 \times
10^{-16}$ erg cm$^{-2}$ s$^{-1}$); afterward the line was not
detected. Other He~I lines seem to have a similar evolution.
  
The strength of the He~I 7065 A\/ line with respect to other lines,
e.g. 5876 A\/, suggests that nonthermal processes are at work.  A
similar evolution of He lines was observed in SN 1996L
\cite{bene99}.
 
\section{\bf Discussion}\label{disc}

The light curves of the type II \g\/ presented in Sect.~\ref{phot} 
show relatively high luminosity  and very slow 
luminosity declines. The absolute magnitude ($M_{V}\sim -18.47$) is 
similar to that of \z\/ and SN~1994W \cite{soll98}, two objects 
sharing some of the properties of \g.

The slow fading of \g\/ persists well beyond 5-6 months, an epoch at
which the light curves of normal SNe II, powered by the radioactive
decay of $^{56}$Co, decline at about 1 mag/100$d$ (Fig.~\ref{abs} and
\ref{homer}).  Therefore, another source of energy in addition to the 
radioactive decay is needed to power the light curve of \g. The
similarity at all epochs \z\/ suggests that also in this case the
surplus energy comes from the transformation of kinetic energy of the
ejecta into radiation owing to interaction with a dense CSM.

Aretxaga et al. \shortcite{aret99} have proposed that in the case of
\z\/ most of the energy have been emitted as X--rays and as ionizing
radiation. They estimated that the total radiated energy was as high
as $10^{52}$ ergs, suggesting complete reprocessing of the mechanical
energy of the ejecta in only a few years. Unfortunately no X--rays
observation is available for \g.  The ionization energy computed from
the
\Ha\/ luminosity \cite{aret99} is considerably
smaller for \g\/ than for \z, because of the weaker line emission
(Fig.~\ref{hal}). However, despite significant differences in the
\Ha\/ luminosity, the bolometric luminosities of the two SNe are very
similar. The radiated energy of \g\/ integrated over the 3 years of
observations is almost the same as in \z\/ in the same period
($10^{50}$ ergs in BVR).

The optical light curves of \z\/, the X--rays and \Ha\/ emission, and
the line--width evolution are reproduced by a model in which the
ejecta is interacting with a dense ($n=10^{7}$ cm$^{-3}$) and
homogeneous CSM (Aretxaga et al., 1999; Terlevich et al., 1992).  In
these conditions the SN remnant evolves very rapidly and radiative
cooling becomes important well before the thermalization of the
ejecta.  The ejecta-CSM interaction causes the formation of a hot
shocked shell between two shock waves: a forward shock ($v \approx$
10000 km s$^{-1}$) encountering and heating the CSM to a temperature
of about $10^{9}$ K and a slow reverse shock ($v \approx$ 1000 \kms)
which thermalizes the SN ejecta to a temperature of about $10^{7}$
K. A high flux of X--rays photons is produced in the high-density
material shock front, while the cooling shocked gas reinforces the
intensity of H$\alpha$ components. The reference model by Terlevich et
al. \shortcite{terl00} of the interaction of the ejecta with a
circumstellar shell ($n_{\circ}=10^7$ cm$^{-3}$ and $Z=10Z_{\odot}$)
does not fit completely the observations of \g. In particular, the
observed \Ha\/ luminosity of \g\/ after 230 days (1 t$_{sg}$) is
fainter by a factor 10 and the Balmer decrement is too steep. Indeed
these differences are not surprising, because type IIn SNe exhibit
considerable heterogeneity, indicating that the physical conditions
and/or the geometry of the interaction can be very different. For
example \z\/ remained visible for over 10 years, while SN~1994W dropped
in luminosity after four months probably because the interaction
ceased earlier.
\cy, the brightest SN~IIn rapidly faded
below the detection limit after two years \cite{tura00}.  

The first spectrum of \g\/ shows a continuum corresponding to
T$_{BB}$=8800 K, a temperature close to that of SN~1987A a couple of
days after the explosion, supporting the idea that \g\/ was discovered
soon after the explosion (Sect.~\ref{lcsect}). On the other hand, the
small temperature variation over the first month (T$_{BB}$(36d)=7000
K) gives the opposite indication. The spectra show some distinctive
feature compared to those of \z, which shared similar bolometric
evolution. The narrow absorption components were never detected in the
spectra of \z, although this might be due to the modest spectral
resolution of those observations.

The spectra of \g\/ are very similar both as line ratios and profiles
to those of \w\/ around maximum \cite{soll98}. In particular, \Ha\/
shows the same velocity profiles with absorption minima blue shifted
by similar amounts (cfr. Fig.~\ref{lineprof} with Fig.~3 in Sollerman
et al., 1998). Other ions, e.g.  Ca~II and Fe~II, show similar line
profiles. Instead the blackbody temperature of \g\/ is lower, and
two days after the discovery it corresponds to that of \w\/ on day 80.

From the blackbody fits and the bolometric luminosity we can derive the
photospheric radii on days 2 and 36, which are of $1.1 \times 10^{15}$ cm
and $1.6 \times 10^{15}$ cm, respectively. 

Up to about 1 yr the P-Cygni profiles of \Ha\/ and Fe lines are
rounded and the narrow emission and absorption components are
symmetric.  This indicates that optical depths of these lines are $>>
1$, the lines are scattering dominated and originate in a rather thin
shell close to a (pseudo)photosphere.  Rybicki and Hummer (1978)
showed that in this situation of an expanding shell the optical depth
in a line depends on the velocity gradient. Fransson(1984) elaborating
on this presented a precise expression for the optical depth which was
then used in the specific case of SN~1994W by Sollerman et
al. \shortcite{soll98} to derive a lower limit to the density in the
shell under the assumption that the lines had a large optical depth
because their profiles were rounded or parabolic \cite{mihalas}. The
case is, in this respect, very similar to that of \g\/ thus we can use the
expression $n_H >> 3 \times 10^8 v_3 r^{-1}_{15}$ cm$^{-3}$, where
$v_3$ is the shell velocity in units of $10^3$ \kms\/ and $r_{15}$ is
the shell radius in units of $10^{15}$ cm.  Assuming that
the shell radius is close to that of the photosphere, using the values
of $v_3$ derived from the minima of the absorptions, we get $n_H >>
10^{8}$ cm$^{-3}$, consistent with the absence of forbidden lines. The
density of the slowly expanding material in \g\/ is therefore similar
to that of SN~1994W but higher than in \z, for which the ratio of
[O~III] 4363, 4959 and 5007 A\/ lines indicate densities
$n_e \geq 10^{7}$ cm$^{-3}$ (Statakis \& Sadler, 1991; Filippenko,
1991).

Another distinctive characteristic of some SNe IIn, e.g. \z, SN~1995N
(Benetti et al., 1995; Turatto et al., 2002) and SN~1986J
\cite{tura93}, is the presence of narrow emission lines of very high
ionization (e.g. [Fe~VII], [Fe~X]) which are direct evidence of energy
dissipation in powerful shock waves. In \g\/ such lines are not
present.  

We noted in Sect.~4 that it is possible to identify gas with
different kinematics. The results of a multicomponent fit for \g\/
have been summarized in Tab.~\ref{gauss}. Compared to the expansion
velocity of the broadest component of \z\/ (FWHM$=15000$ \kms), the
broad component in \g\/ is much slower (FWHM$<5000$ \kms), while the
other components are similar. This implies that the fast--moving
ejecta photoionized by the hard radiation of the shock move more
slowly, justifying the absence of very high ionization species.

Similarly to SN~1996L \cite{bene99}, after day 200 high excitation He
lines emerge. The low temperature of the envelope suggests that these
lines result from non-thermal processes.  The unusual relative
strengths of the He~I lines probably owe their origin to excitation by
a radiative shock in the dense or clumpy wind with the consequent
large optical depth in the He~I 3889A line. Such line ratios have been
observed in some symbiotic stars by Proga et al. \shortcite{proga} who
appealed to the calculations of Almog and Netzer \shortcite{almog} to
show that large optical depths of He~I 3889A combined with high
densities $10^8 - 10^{10}$ may produce these unusual line ratios. We
have demonstrated the plausible existence of high densities at
intermediate velocity probably associated with shock propagation. The
bump in the \Ha\/ light curve during the same interval may result from
the same shock excitation.

After 2 years, the absorption components disappear and \Ha\/ evolves to a
boxy profile. Simultaneously the \Ha\/ luminosity declines and the light
curves become steeper, but still not as much as the decay rate of $^{56}$Co. 
This may suggest that the interaction is fading.  Something
similar happened in SNe 1994aj \cite{bene98} and 1996L \cite{bene99}
in which the narrow P-Cygni profiles of gradually
disappeared, leaving broad boxy emissions.
In the case of SN~1996L we observed the
progressive development of the narrow features first as pure
emissions, and later as P--Cygni structures, when significant 
recombination took place. After 100 days the lines broadened and
the absorption progressively disappeared. This has been interpreted as the
onset of the interaction of the ejecta with an outer shell which is
progressively swept away. The analysis of the timing of these phases
led to the conclusion that the shell was produced by a
wind episode that started 9 years before the explosion and lasted 6 years.
Benetti et al. \shortcite{bene99} noticed that SNe 1994aj and 1996L
had linear light curves in the early 100 days, which is indirect 
evidence of low--mass, wind--deprived envelopes.

The case of \g\/ is different. The light curve remains very flat since
the discovery, indicating that the interaction with the CSM started
soon after the burst, and is still continuing at the time of our last
observation on day 942. Therefore, if the CSM is due to a stellar wind
from the progenitor, a strong wind persisted almost to the time of the
explosion.  From the duration of the interaction, given a wind
velocity is 750 \kms\/ and a maximum expansion velocity of the ejecta
of 4500 \kms (Fig.~\ref{lineprof}), we derive that the wind episode
started at least a dozen years before the burst. The dense CSM could
be due also to the presence of a companion which might have stripped
gas from the precursor or might have lost its own envelope.

The fact that the P-Cygni profile is already well developed at the
discovery, contrary to the case of SN~1996L, indicates that the
recombination of the shell took place earlier, probably because of
higher densities ($n_H >> 10^{8}$ cm$^{-3}$). After 2 years, however,
most of the slowly expanding shell has been swept away by the forward
shock, the optical depth of the unshocked material is low, and the
line profiles show only the boxy shape due to the shock. The
simultaneous decrease of brightness (increase of decline rate of the
light curves) can be evidence of a steepening of the density profile
in CSM shell.  Yet, up to 3 years after the discovery, there is
evidence that the residual interaction between the SN ejecta and CSM
has not completely ceased.

Thus, \g\/ seems to be an intermediate case between
\z\/ interacting with a dense CSM from the early stages after the
explosion, and less extreme objects like SNe 1994aj and 1996L which
interaction with a shell of material ejected a few years before the
explosion were limited in time.

\subsection{\bf Progenitor Mass Loss}\label{mass}

In Sect.~\ref{disc} we pointed out that the Terlevich et al.
\shortcite{terl00} model of the interaction of the ejecta with a dense,
uniform density gas, which reproduces the observations of \z, does not
fit all the observables of \g. Chugai \& Danziger
\shortcite{chug94} presented calculations of the dynamical interaction
of the ejecta of \z\/ with a dense clumpy circumstellar wind ($\rho
\propto r^{-2}$).  Following the latter approach we can derive some
important parameters of the progenitor.

We assume that the velocity of the broad H$\alpha$ emission is
representative of the velocity of the ejecta, since this component is
supposed to originate in the shocked ejecta.
 
The interaction between the ejecta and the progenitor's wind follows
two distinct phases: initially the outer part of the SN ejecta, having
a power-law density distribution, collides with the wind.  This is the
so-called free expansion phase. In this phase the shock velocity
(which is observed as the velocity of the broad component of
H$\alpha$) is given by

\begin{equation} 
 v = 7.46 \times 10^8 E_{51}^{2/5} 
  \left(\frac{M_{ej}}{M_{\odot}}\right)^{-1/5} w_{16}^{-1/5} t_{yr}^{-1/5} [{\rm cm~s}^{-1}], 
\end{equation}

\noindent
where $E_{51}$ is the kinetic energy of the ejecta in units of $10^{51}$erg, 
$t_{yr}$ is the time elapsed since the explosion in years, 
$w=\dot{M}/u_w$ is the wind density parameter, $u_w$ is the wind velocity
and $w_{16}=w/10^{16}$ g cm$^{-1}$. A lower limit for $w$ can be obtained from
the general expression for the luminosity of the shock wave: 

\begin{equation} 
 L = \frac{1}{2} \psi w v^3. 
\end{equation} 

\noindent
Here $L$ is the SN luminosity (which is dominated by the interaction) and
$\psi$ is the efficiency of conversion of mechanical into optical energy.

In the case of SN~1995G at $t \approx 0.2$ years, we estimate 
$L \approx 4 \times 10^{42}$erg s$^{-1}$ (with a considerable uncertainty, 
because of the paucity of observed data at this epoch) and 
$v \approx 4000$ km s$^{-1}$. If we assume $\psi=1$, equation (2)
gives $w_{16} = 12.5$ g cm$^{-1}$. For a wind velocity $u_w = 10$ km s$^{-1}$,
typical of a red supergiant (RSG), we get 
$\dot{M} = 0.002 M_{\odot}$ yr$^{-1}$, 
which is a lower limit because in general $\psi<1$. Such large $\dot{M}$ 
results from the fact that $L$ is large but $v$ is small. 
In the case of SN~1988Z, which has the same $L$ but $v = 17000$ km s$^{-1}$, 
$\dot{M} = 2.6 \times 10^{-5} M_{\odot}$ yr$^{-1}$. 
From Eq. (1) and $w_{16} = 12.5$ g cm$^{-1}$ we obtain the ejecta mass of 
SN~1995G: $M_{ej}/M_{\odot} \simeq 9 \times E_{51}^2$. So, if
$E \simeq 10^{51}$erg, $M_{ej} \simeq  9M_{\odot}$.

Eventually, the outer, power--law part of the ejecta overtake the wind, and the
interaction is then between the wind and the inner part of the ejecta, which
has a flat density profile. In this regime, called the blast wave phase 
\cite{chug94}, the velocity is 

\begin{equation} 
 v = (2 \alpha E/(M + w r))^{1/2},
\end{equation} 

\noindent
where $E$ and $M$ are the kinetic energy and the mass of the ejecta,
respectively, and $\alpha$ is the ratio of kinetic to total energy 
(typically 1/2). For the parameters of SN~1995G, this gives 
$M_{ej}/M_{\odot} = 3 E_{51}$, so if $E \simeq 10^{51}$erg, 
$M_{ej} \simeq 3M_{\odot}$.

Since it is not clear which phase of the interaction was observed in
SN~1995G, the mass of the ejecta can be estimated from the above
considerations to be in the range $3-9 M_{\odot}$ provided that
$\psi=1$ and $E = 10^{51}$erg. 
Including the mass of the
neutron star remnant, we obtain a pre--SN mass between 5 to 11
$M_{\odot}$, which correspond to zero--age main sequence masses around
$15-20M_{\odot}$ if a new empirical formulation of the mass--loss rate
and a diffusive approach of turbulent mixing is adopted \cite{sala99}.
We stress that the values of the mass loss
are very sensitive to metallicity and that the estimates  of $M_{ej}$
are strongly dependent on the value of the explosion energy.
The uncertainties on the progenitor mass are therefore very large,
meaning that it could be significantly lower.

As for the H$\alpha$ luminosity, this is a constant factor of 6
smaller in SN~1995G than in SN~1988Z. However, the width of the
intermediate component of H$\alpha$ is about a factor of 4-5
smaller (2000 vs. 9000 km s$^{-1}$), so if the wind is clumpy as in
SN~1988Z, but the density is higher in SN~1995G ($\dot{M}$ was
higher), this may be a natural result.

We note also that the small inferred mass of the precursor of \z\/
($8-10M_{\odot}$) is a consequence of the assumptions $E = 10^{51}$erg and
$\psi=1$. If a higher value of the kinetic energy is used for SN~1988Z, as may
be suggested in Aretxaga et al. \shortcite{aret99}, a larger ejecta mass is
obtained.  On the other hand decreasing $\psi$ leads to a higher wind density
and mass--loss rate (and to a larger mass lost in the wind), but it also leads
to a smaller ejecta mass through eq. (1). Whether this leads to a
significantly different estimate of the mass of the progenitor is not clear,
however, as this depends on the duration of the strong RSG wind episode.

\section{\bf Conclusions}

This paper presents the photometric and spectroscopic observations of \g\/
obtained at ESO La Silla, Lick and Asiago, over a period of more than 4 years 
after the discovery.

The broad band and bolometric light curves are flatter than those
expected from radioactive decay. The luminosity peak and evolution are
very similar to those of the well studied \z. This implies the
presence of an additional source of energy, most likely the
interaction of the ejecta with a dense CSM.

The line profiles show evidence of different components. The strongest
line, \Ha, shows relatively narrow P-Cygni profiles with minima
displaced by 700-800 \kms superimposed on an intermediate width
emission FWHM$ \approx2000$ \kms and on a broader component having
FWHM$ \approx4000$ \kms. The widths of these emission lines decrease
with time.  At 942 days after maximum the \Ha emission has a boxy shape with
FWZI=2500 \kms, suggesting that most of the expanding shell has been
swept away and the optical depth of the material is low.  The emission
lines of \g\/ were narrower than in \z, indicating slower expansion
velocities of the ejecta.

The detection of strong He lines between about 200 and 600 days after the 
discovery indicates the presence of non-thermal effects. 
Simple considerations about the density of the slowly expanding material 
indicate densities $n_H >> 10 \time 10^8$ cm$^{-3}$, in agreement
with the absence of forbidden lines.

In the hypothesis that the additional source of luminosity of \g\/
is the conversion to radiation of the kinetic energy of the ejecta due
to interaction with a precursor wind, we estimate the mass of the ejecta and 
the mass loss.
We obtain $M_{ej} \simeq 3-9$ $M_{\odot}$ and $\dot{M}$ = 0.002
$M_{\odot}$/yr, respectively, provided that the conversion efficiency
is $\psi=1$ and the explosion energy is $E = 10^{51}$erg.

\section*{Acknowledgments}
We acknowledge support from the Italian Ministry for
University and Scientific and Technological Research (MURST) through grant
Cofin MM02905817.\\
A.V. Filippenko is grateful for a Guggenheim Fellowship
and for NSF grant AST--9987438.

\end{document}